\documentclass[prd,superscriptaddress,twocolumn,nopreprintnumbers,floatfix]{revtex4}
\usepackage{graphicx,url,amssymb,amsmath,longtable,rotating,color,units,wasysym,subfigure,epsfig,multirow,amsbsy,soul,mathrsfs,amsmath,xcolor,mathtools}
\usepackage[plainpages=false, colorlinks=true, anchorcolor=blue, linkcolor=blue, citecolor=blue, bookmarks=false]{hyperref}
\usepackage{float}
\usepackage{resizegather}
\usepackage{doi}
\renewcommand{\o}{\varnothing}
\newcommand{\SPA}{School of Physics and Astronomy, Monash University, Clayton VIC 3800, Australia}
\newcommand{\OzGravMonash}{OzGrav: The ARC Centre of Excellence for Gravitational Wave Discovery, Clayton VIC 3800, Australia}

\begin{document}

\title{Gravitational-wave astronomy with a physical calibration model}
\author{Ethan Payne}
\email{ethan.payne@ligo.org}
\affiliation{\SPA}
\affiliation{\OzGravMonash}

\author{Colm Talbot}
\affiliation{LIGO, California Institute of Technology, Pasadena, CA 91125, USA}
\affiliation{\SPA}
\affiliation{\OzGravMonash}

\author{Paul D. Lasky}
\author{Eric Thrane}
\affiliation{\SPA}
\affiliation{\OzGravMonash}

\author{Jeffrey S. Kissel}
\affiliation{LIGO Hanford Observatory, Richland, WA 99352, USA}

\begin{abstract}
We carry out astrophysical inference for compact binary merger events in LIGO-Virgo's first gravitational-wave transient catalog (GWTC-1) using a physically motivated calibration model.
We demonstrate that importance sampling can be used to reduce the cost of what would otherwise be a computationally challenging analysis.
We show that including the physical estimate for the calibration error distribution has negligible impact on the inference of parameters for the events in GWTC-1.
Studying a simulated signal with matched filter signal-to-noise ratio $\text{SNR}=200$, we project that a calibration error estimate typical of GWTC-1 is likely to be negligible for the current generation of gravitational-wave detectors.
We argue that other sources of systematic error---from waveforms, prior distributions, and noise modelling---are likely to be more important.
Finally, using the events in GWTC-1 as standard sirens, we infer an astrophysically-informed improvement on the estimate of the calibration error in the LIGO interferometers.
\end{abstract}

\maketitle
\section{Introduction}
The burgeoning field of gravitational-wave astronomy \cite{GW150914, GW170817,gwtc-1} is advancing our understanding in multiple fields of astrophysics, including cosmology \cite{hubbleGW}, galactic and stellar evolution \cite{gwtc1pop}, and strong-field gravity \cite{gwtc1GR}.
As the sensitivity of observatories improve, and gravitational waves are observed with increasingly large signal-to-noise ratios (SNRs) \cite{schutz2011networks, vitale2016, ligo2019gravitational}, an understanding of systematic effects will become ever more important.
Sources of systematic biases include errors associated with gravitational waveforms~\cite{Purrer2020}, imperfect prior distributions, incorrect estimates for the noise power-spectral density \cite{sylviaPSD, student-t, chatziioannou2019noise}, and errors associated with the calibration of the detectors~\cite{Vitale2014}.
Here, we focus on errors associated with calibration.

Calibration is defined as the process of converting the detector's primary control system error signal due to differences in the lengths of the interferometer's arms to an estimate of strain on the detector \cite{abbott2017calibration, Cahillane2017, acernese2018calibration, Sun2020}.
Imperfect knowledge of the interferometer's control system and response to diffential arm length changes leads to systematic error in the amplitude and phase of the calibration. This error is estimated by conducting a vast suite measurements of the control system, and propagating the results of those measurements into a physically informed model. The resulting error estimation is represented by a frequency-dependent probability distribution.
In order to avoid bias, the estimated probability distribution of calibration errors must be taken into account when inferring the astrophysical parameters of gravitational-wave signals~\cite{Vitale2014}.
Unfortunately, marginalizing over calibration error distributions can dramatically increase the number of parameters used in astrophysical inference: from the 15 required to describe a binary black hole to $> 50$ \cite{spline_doc}.
This increase in parameter space can lead to a significant increase in computational cost and convergence issues, which has somewhat limited efforts to carry out astrophysical inference that include an estimate of calibration errors up to this point.

In this work, we demonstrate a computationally efficient implementation of the original physical calibration model \cite[][see Sec.~\ref{sec:model}]{Cahillane2017,Sun2020} for astrophysical inference.
Following~\cite{hom}, we first evaluate the posterior distribution of astrophysical parameters without any estimated calibration error distribution, then employ importance sampling to reweight approximate results to include this contribution. 
Importance sampling~\cite{Robert&Casella,Liu} is the technique of constructing weights for individual samples which determine each sample's contribution to the inferred probability distribution.
Having verified the analysis procedure, we carry out a study of gravitational-wave signals from the first LIGO-Virgo Gravitational-Wave Transient Catalog (GWTC-1)~\cite{gwtc-1} using estimates of the calibration error at the time of those events.
Combining data from multiple events, we infer an astrophysically informed calibration error estimate~\cite{Essick2019}, showing that it is possible to learn about the Advanced LIGO interferometers~\cite{aLIGO} using gravitational waves as standard sirens.

With the assumption that the estimated systematic error present in GWTC-1 will be typical in the future, we demonstrate the effect of the calibration error estimate on a simulation of an $\text{SNR}=200$ binary black hole merger ---approximately the loudest event that will be observed by the second-generation interferometer network during its operational lifetime \cite{schutz2011networks, chen2014loudest}.
In this regime, naive application of our importance sampling algorithm becomes inefficient.
However, we find that the calibration error estimate still has only a marginal effect on the posterior distributions of the intrinsic astrophysical parameters. 
The localization parameters are the most affected by the inclusion of the calibration error distribution; the sky map credible region approximately doubles in size for a $\text{SNR}=200$ event.
We conclude that the impact of calibration error estimates will likely be small compared to other previously mentioned sources of systematic error in astrophysical parameter estimation.

The remainder of this paper is organized as follows.
In Section~\ref{sec:model}, we summarize the physically motivated calibration model, its parameters, and the collective error estimate from ~\cite{Cahillane2017,Sun2020}.
In Section~\ref{sec:method}, we describe our methodology for efficiently marginalizing over the probability distribution of calibration errors with importance sampling.
In Section~\ref{sec:testing}, we demonstrate our implementation using simulated data while in Section~\ref{sec:results}, we analyze data from GWTC-1.
We end with concluding remarks in Section~\ref{sec:conclusions}.

\section{Calibration model}\label{sec:model}
In this section, we summarize the physical calibration model for the LIGO detectors described in~\cite{Cahillane2017,Sun2020}.
Though the Virgo detector~\cite{acernese2014advanced} is similar to the LIGO detectors, the systematic error probability distribution used in this study is informed by the 68\% confidence interval bounds on the systematic error described in Ref.~\cite{acernese2018calibration}.

\subsection{The physical model}
The LIGO detectors are dual-recycled, kilometer-scale Fabry-P\'erot Michelson interferometers, most sensitive between 10 and 2000 Hz~\cite{aLIGO}. 
The passage of a gravitational-wave signal induces differential displacement between the two arms of the interferometer. 
This differential arm displacement is measured at the output of the detector, where interfering laser light reflected from the resonant arm cavities is incident on a set of photodiodes~\cite{Izumi2016}. 
The photodiode signals are summed and digitized to form a signal which includes both detector noise and gravitational waves. 
The digitized signal also serves as the residual error signal of the feedback control system for the changes in differential arm length. 
This, among other control loops in the detector, ensures that external noise sources do not force the interferometer cavities off-resonance. This is achieved through differentially actuating on the arm cavity mirrors, or test masses, and their suspension systems~\cite{robertson2002quadruple, aston2012update,carbone2012sensors}. 
Below $\sim100$\,Hz, the control systems actuation forces suppress the interferometer's response to differential arm displacement. Above this frequency, the response is free of control system influence and depends on both the interferometric response to differential arm displacement as well as the signal processing electronics of the photodiodes. 
Thus, in order to reconstruct the measured strain from the digitized photodiode signal over the entire sensitive frequency band, a physically motivated model of the response and control system are required.

The model is divided into two conceptual components. The first component, the \textit{sensing function}, is an optomechanical description of the interferometer if it were free of control forces, photodiode signal processing electronics and the digital acquisition system. 
The second component, the \textit{actuation function}, describes how the control system splits the single, digitally filtered, error signal among three stages of cascading actuators on the test mass quadruple suspension systems; incorporating those actuators' digital to analog converters and signal processing electronics. 
The actuation function also includes the complex displacement response of the test mass for those forces from each stage \cite{robertson2002quadruple, aston2012update,carbone2012sensors}.
Both model components are frequency-dependent complex transfer functions that are mostly static in time, but each have slowly time-varying correction factor parameters to account for natural drifts in their behavior.

The calibrated strain signal $h$ is related to the differential arm error signal $d_\text{err}$ (in the frequency-domain) by the response function $R$,
\begin{eqnarray}\label{eq:calibration}
h & = &  \frac{1}{L}\,R\,d_\text{err} ~=~ \frac{1}{L} \left(\frac{1 + CDA}{C}\right)
    d_\text{err} .
\end{eqnarray}
Here, $C$ is the sensing function while $A$ is the actuation function.
The variable $D$ describes the set of digital filters responsible for converting the error signal to the single differential arm control signal.
The variable $L$ is the length of the interferometer arms.
Equation~\eqref{eq:calibration} illustrates how systematic errors in $C$ and $A$ lead to an error in $h$.

Following~\cite{Cahillane2017,Sun2020}, we employ the following model for the sensing function:
\begin{align}\label{eq:sensing}
    C(f|\Lambda)= & 
    \frac{\kappa_C(t) H_C}{1+if/f_{CC}(t)} C_R(f) 
    e^{-2\pi i f \tau_C} \nonumber\\
    & \times \frac{f^2}{f^2 + f_S^2 - i f f_S/Q} .
\end{align}
Here, $f$ is frequency, and $\{H_C,f_{CC}, \tau_C, f_S, Q\}$ are the parameters describing the optomechanical response (summarized in Table~\ref{tab:sensing}), which are part of a larger set of calibration model parameters $\Lambda$.
The parameters $\kappa_C(t)$ and $f_{CC}(t)$ represent the time-dependent corrections needed to account for alignment drift in the suspended cavities.
Detuning between the signal recycling cavity  and arm cavities~\cite{miyakawa2006measurement} is modelled as an optical spring with a characteristic frequency, $f_S$, and associated quality factor, $Q$. 
Finally, $C_R(f)$ is the digital acquisition response, which is measured a priori with high-precision. 
The probability distributions for the time-independent parameters of the sensing function, $\Lambda$, are determined by fitting measurements from a single, representative reference time with the model outlined in Eq. \eqref{eq:sensing} using Markov Chain Monte Carlo (MCMC) sampling~\cite{emcee}. 
The probability distribution for the parameter $f_{CC}$ is determined both by an MCMC fit from the reference time measurement, and, like $\kappa_{C}(t)$, by the continuous high-precision tracking of its time dependence~\footnote{The analysis undertaken in this manuscript does not incorporate the probability distribution from MCMC sampling for $f_{CC}(t)$ following an implementation error in the analysis undertaken in Ref.~\cite{Sun2020}. This omission has only a marginal effect on the calibration uncertainty estimate.}.

\begin{table}[t]
    \centering
    \begin{tabular}{|c|c|c|}
        \hline
        symbol & name & units \\
        \hline\hline
        $H_C$ & optical gain & $\unit{cts/m}$ \\\hline
        $f_{CC}$ & coupled cavity pole frequency & $\unit{Hz}$ \\\hline
        $\tau_C$ & time delay & $\unit{\mu s}$ \\\hline
        $f_S$ & optical spring frequency & $\unit{Hz}$ \\\hline
        $Q$ & optical spring quality factor & --- \\\hline
    \end{tabular}
    \caption{Sensing parameters. The optical gain describes the overall magnitude of the sensing function. The coupled cavity pole frequency describes the bandwidth of the interferometer arm and signal recycling cavity system. The time delay compensates for light travel time within the arms and computational delay in the photodiode analog-to-digital conversion process. The optical spring parameters describe the characteristic frequency and amount of detuning between the arm and signal-recycling cavities.}
    \label{tab:sensing}
\end{table}

The actuation function is modeled as follows:
\begin{align}
    A(f|\Lambda) = & \left.\kappa_U(t) F_U(f) H_U A_U(f) e^{-2\pi i f \tau_U} + \right. \nonumber\\
    & \left.\kappa_{P}(t)F_P(f) H_P A_P(f) e^{-2\pi i f \tau_P} + \right. \nonumber\\
    & \left. \kappa_{T}(t)F_T(f) H_T A_T(f) e^{-2\pi i f \tau_T}. \right. 
\end{align}
Here, $\{H_U, \tau_U, H_P, \tau_P, H_T, \tau_T\}$ are the actuation calibration parameters summarized in Table~\ref{tab:actuation}.
The subscripts refer to the stage of the suspension system where actuation force is applied.
These are the ``upper-intermediate'', $U$, ``penultimate'', $P$ and ``test mass'', $T$.
The force-to-displacement response and the response of actuator electronics are incorporated in $A_i(f)$. 
The digital distribution filters, $F_i(f)$, and the scalar time-varying correction factors, $\kappa_i(t)$, are precisely known, and so do not appear in Table~\ref{tab:actuation}.
Again, the prior distributions for the actuation parameters are determined by MCMC sampling with data from single measurements of each stages' response. 
The values and uncertainties associated with time-dependent quantities are computed at a 1\,\textrm{hr} cadence over the duration of an observing run. 

\begin{table}[t]
    \centering
    \begin{tabular}{|c|c|c|}
        \hline
        symbol & name & units \\
        \hline\hline
        $H_U$ & upper intermediate stage gain & $\unit{N/cts}$ \\\hline
        $\tau_U$ & upper intermediate stage delay & $\unit{\mu s}$ \\\hline
        $H_P$ & penultimate stage gain & $\unit{N/cts}$ \\\hline
        $\tau_P$ & penultimate stage delay & $\unit{\mu s}$ \\\hline
        $H_T$ & test mass stage gain & $\unit{N/cts}$ \\\hline
        $\tau_U$ & test mass stage delay & $\unit{\mu s}$\\\hline
    \end{tabular}
    \caption{Actuation parameters. The scalar gains applied to each of the stages calibrates the overall magnitude of the actuation. The time delays arise from computational delays in the digital-to-analog system.}
    \label{tab:actuation}
\end{table}

The final parameter within the physical calibration model is an overall scalar magnitude factor, $\eta_\textrm{PCAL}$, whose probability distribution is derived from any systematic error and uncertainty in the photon calibrator systems (PCALs). The photon calibrator systems are used as fiducial displacement references for each detector \cite{karki2016advanced,bhattacharjee2020fiducial}.
Typically, the systematic error is negligibly different from unity, and only adds an overall magnitude uncertainty: coincidentally $0.79\%$ for both LIGO detectors during the second observing run. 
This additional correction is applied as a multiplicative factor to the response function~\footnote{While each detectors fiducial reference has its own systematic error and associated uncertainty, the collection of references for entire network are seeded from a single global reference. This common error on the global reference is included in the uncertainty for each detector, and excluded as an independent error from this analysis as it is degenerate with the luminosity distance of a gravitational-wave source~\cite{bhattacharjee2020fiducial}.}.

\subsection{The phenomenological model}
While the physical model of the response function, $R(\Lambda)$, produces an approximately correct response, inspection of an ensemble of frequency-dependent residuals $R_\text{measured}/R(\Lambda)$, constructed from sensing and actuation function measurements, shows that the model is incomplete, i.e., the residuals are not consistent with unity; see Fig.~11 from~\cite{Sun2020}.
The authors of~\cite{Sun2020} build an additional phenomenological model for $C$ and each stage of $A$ on top of the physical model in order to estimate the residuals, completing the error estimate with new phenomenological parameters.

The phenomenological model employs Gaussian process regression of the residuals, interpolating between 128 frequency points \cite{gpr, scikit_gpr} and several hyper-parameters constraining the covariance kernel between each frequency point.
The corrected sensing and actuation functions are given by
\begin{align}
    C'(\Lambda) = & \eta_C(\Lambda) C(\Lambda), \\
    A'(\Lambda) = & \eta_A(\Lambda) A(\Lambda) .
\end{align}
Here, $(C,A)$ are the physical sensing and actuation models while $(C',A')$ are the phenomenologically-corrected models. They are included as a part of the frequency-dependent estimated distribution of calibration error.
Since $\eta_C$ and $\eta_A$ for each stage are complex-valued functions described by a magnitude and phase, the phenomenological model introduces an additional $256\times4$ calibration parameters to $\Lambda$.

After applying both physical and phenomenological models to LIGO data, the authors of~\cite{Sun2020} find that the distribution of errors in the response $R$ completely explains $R_{\textrm{measured}}/R(\Lambda)$, and is dominated---in most frequency regions---by uncertainty from the Gaussian process fit.
That is, the systematic error from imperfect design of the physical model is large compared to the uncertainty in its parameters.
However, by introducing such a high-dimensional phenomenological model, the systematic error of the physical model is converted almost entirely into statistical uncertainty.
With so many free parameters, we expect it should be possible to fit nearly any measured form of $R$.

\section{Method}\label{sec:method}
Our goal is to estimate astrophysical parameters $\theta$ describing the gravitational waveform of a compact binary merger given strain data $h$ and marginalizing over the unknown calibration parameters, $\Lambda$.
We follow style conventions from~\cite{intro}.
Assuming Gaussian noise, the likelihood is given by:
\begin{align}\label{eq:bins}
    {\cal L}(h_j|\theta,\Lambda) = \frac{1}{2\pi P_j}
    \exp\left(-2\Delta f \frac{\left|h_j-\lambda_j(\Lambda)\mu_j(\theta)\right|^2}{P_j}\right).
\end{align}
Here, $P_j$ is the power spectral density of the interferometer, $\mu_j(\theta)$ denotes the gravitational-wave model. 
In this manuscript, we utilize \textsc{IMRPhenomPv2}~\cite{imrphenompv2_1, imrphenompv2_2} for our source model of binary black hole systems, and \textsc{IMRPhenomPv2NRTidal}~\cite{dietrich2017closed} for binary neutron star mergers. 
The parameters of the compact binary coalescence, $\theta$, include intrinsic properties such as the masses and spins of the individual compact objects, and extrinsic parameters informing the orientation and location of the binary system. 
The subscript $j$ refers to a single frequency bin, which are spaced by $\Delta f$.
Since the noise in each bin is approximately independent, the combined likelihood is simply
\begin{align}\label{eq:prod}
    {\cal L}(h|\theta,\Lambda) = \prod_j {\cal L}(h_j|\theta,\Lambda) .
\end{align}
The product over frequency bins is implied in subsequent equations.
Meanwhile, the calibration error is described by
\begin{align}
    \lambda(\Lambda) = \frac{R(\Lambda)}{R_{\o}} ,
\end{align}
the ratio of the true response function $R(\Lambda)$, which depends on calibration parameters $\Lambda$, to the original response function used to calibrate the data $R_{\o}$, denoted as $\eta_R$ in Ref.~\cite{Sun2020}.

Our \textit{target distribution}, the one for which we want to generate posterior samples, is Eq.~\eqref{eq:prod} marginalized over $\Lambda$:
\begin{align}\label{eq:marginalised}
    {\cal L}_\Lambda(h|\theta) = \int d\Lambda \,
    {\cal L}(h|\theta,\Lambda) \pi(\Lambda) .
\end{align}
Here $\pi(\Lambda)$ is our prior on the calibration parameters.
The target distribution can be computationally expensive to sample from owing to the extra dimensionality associated with $\Lambda$.
However, if the original calibration $R_{\o}$ is at least approximately correct, and if the SNR of the event is not too large (we quantify how large momentarily), then we can employ importance sampling to avoid sampling in $\Lambda$.

Following~\cite{hom}, we define our \textit{proposal distribution},
\begin{align}
    {\cal L}_{\o}(h_j|\theta) = \frac{1}{2\pi P_j}
    \exp\left(-2\Delta f \frac{\left|h_j-\mu_j(\theta)\right|^2}{P_j}\right) ,
\end{align}
corresponding to the likelihood we would use if we believed the original response function $R_{\o}$ was perfectly accurate.
We use the proposal distribution to generate samples in $\theta$ using the \textsc{Bilby}~\cite{bilby,bilby_gwtc1} implementation of \textsc{Dynesty}~\cite{Dynesty}, a nested sampling algorithm~\cite{Skilling}.
Since we are not sampling in $\Lambda$, the proposal samples are computationally cheap to generate.

Next, for each posterior sample of the binary model parameters, drawn from the proposal distribution, $\theta_i$, we calculate a weight, which requires marginalizing over calibration parameters.
Following~\cite{Sun2020}, we carry out this calculation using a predetermined set of $N=10^4$ calibration response curves, generated with random draws from the prior distribution $\{\Lambda_k\} \sim \pi(\Lambda)$.
We define a doubly-indexed weight relating the proposal likelihood to the target likelihood:
\begin{align}
    w_{ik} = \frac{{\cal L}(h|\theta_i, \Lambda_{k})}{{\cal L}_{\o}(h|\theta_i)} .
\end{align}
Here, $i$ indexes binary posterior samples for the parameter $\theta$ while $k$ indexes calibration prior samples for $\Lambda$.
The calibration-marginalized weight is simply
\begin{align}
    w_{i} &= \frac{1}{N} \sum^{N}_{k} w_{ik} = \frac{{\cal L}_{\Lambda}(h|\theta_i)}{{\cal L}_{\o}(h|\theta_i)} .
\end{align}
Alternatively, we can marginalize over the gravitational-wave model parameters in order to obtain weights useful for constructing posteriors for the calibration parameters:
\begin{align}
    w_{k} &= \frac{1}{n} \sum^{n}_{i} w_{ik} =  \frac{{\cal L}_\theta(h|\Lambda_k)}{{\cal Z}_{\o}(h)} ,
\end{align}
where ${\cal Z}_{\o}(h)$ is the normalization coefficient of the proposal posterior distribution, known as the Bayesian evidence.
This procedure is similar to approaches for estimating neutron-star equations of state with Gaussian processes~\cite{landryEOS}. 

The weights quantify the relative importance of each sample in light of the fact that we are actually interested in the target distribution, not the proposal distribution.
The weights can be input directly into routines for constructing corner plots.
They may also be used to calculate the Bayesian evidence for the target distribution, ${\cal Z}_\Lambda(h)$, from the evidence for the proposal distribution. 
The ratio of the two evidences is simply the average weight, 
\begin{equation}
    {\cal B}^\Lambda_{\o} = \overline{w} = \frac{{\cal Z}_\Lambda(h)}{{\cal Z}_\o(h)},
\end{equation}
known as the Bayes factor which provides a measure of the preference for the calibration model in comparison to the null hypothesis $\o$ that the data are already correctly calibrated. 
The process of constructing these weights is known as importance sampling~\cite{Robert&Casella,Liu}. 
This approach is not confined to the calibration model outlined in Section~\ref{sec:model}, and allows for the application of improved models in the future. 
Furthermore, the method can equivalently be applied with other spline models \cite{spline_doc, Vitale2014} used for analyses in GWTC-1~\cite{gwtc-1}. 

The efficacy of importance sampling can be measured using an efficiency~\cite{Kish, elvira, hom}:
\begin{align}\label{eq:efficiency}
    \epsilon = & \frac{n_\text{eff}}{n} 
    =  \frac{1}{n} \frac{\left(\sum_i^n w_i\right)^2}{\sum_i^n w_i^2} .
\end{align}
Here, $n$ is the number of astrophysical samples generated using the proposal distribution while $n_\text{eff}<n$ is the number of effective samples created from importance sampling.
If the proposal distribution is close to the target distribution, the efficiency will be high.
As a rule of thumb, $\epsilon>50\%$ is ``excellent'' (providing a fast, reliable answer) while $\epsilon\approx1\%-50\%$ is ``good,'' providing adequate efficiency to make importance sampling clearly useful.
Efficiencies $\lesssim1\%$ indicate that the proposal distribution is not necessarily a good approximation for the target distribution, and so reweighting begins to become inefficient, requiring a large number of initial samples and many evaluations of the target likelihood in order to obtain a reliable answer.
The efficiency falls with increasing SNR, since louder events are characterized by progressively peaked likelihood functions. 
We verify that the efficiency is above 10\% when $\textrm{SNR}\lesssim 40$.
One can judge the convergence of the importance sampled result by considering the number of effective samples.
The efficiency can also be used as a measure of the overall effect of the inclusion of a physical calibration model, though there are better measures.
Pathological cases, where importance sampling fails due to multi-modality, are unlikely to apply to our present application; see~\cite{hom} for additional details.

One benefit of likelihood reweighting is its low computational cost.
By directly executing Bayesian inference with the calibration-marginalized likelihood, the number of evaluations of the more computationally expensive model is orders of magnitude larger than the number of posterior samples produced. 
By utilizing likelihood reweighting, the proposal distribution is found with a cheap likelihood function before the expensive likelihood is used sparingly in post-processing.

We can also use the astrophysical parameter-marginalized weights to construct posterior distributions for the calibration hyper-parameters informed by an ensemble of events. We construct weights for the $k$\textsuperscript{th} set of calibration curves informed by $M$ events as
\begin{equation}\label{eq:stacking}
    w^{\textrm{tot}}_k = \prod_\nu^Mw_k^\nu = \prod_\nu^M \frac{{\cal L}_\theta(h^\nu|\Lambda_k)}{{\cal Z}_{\o}(h^\nu)},
\end{equation}
where $\nu$ indexes the different events, not to be confused with the additional implied product over frequency bins in Eq.~\eqref{eq:bins}.
The average combined weight, $\overline{w}^{\textrm{tot}}$, is the Bayes factor for the calibration error distribution compared to the null hypothesis that the calibration error is zero. 
Of course, in order to combine multiple events, we must take care to ensure that the interferometer is in the same state.
Otherwise, the calibration parameters can be different for different events.
Thus, one must ensure that events are only combined for a period during which the interferometer is maintained in a steady configuration.

\section{Simulated Events}\label{sec:testing}
We validate our method using simulated signals injected into Gaussian noise colored to match the Advanced LIGO design sensitivity noise curve~\cite{aLIGO}.
We analyze two signals, both with properties consistent with GW150914~\cite{GW150914}.
We focus on high-SNR events where calibration uncertainty is relatively more important.
In one case, we adjust the distance to achieve an optimal $\text{SNR}=30$, which is comparable to the loudest observed gravitational-wave signal, GW170817~\cite{GW170817} with $\text{SNR}\approx32$.
In the second case, we set the distance to achieve $\text{SNR}=200$.
We use calibration envelopes equivalent to the calibration estimate at the time of GW170817~\footnote{The Virgo observatory uses a spline-based calibration uncertainty model~\cite{spline_doc}.}.

Starting with the $\text{SNR}=30$ event, we compare the posterior distributions for binary parameters $\theta$ obtained three different ways: ignoring calibration error, marginalizing over calibration error estimates with the importance sampling method described above, and with ``direct sampling,'' in which we marginalize over calibration with the $N=10^4$ response curves at every step using Eq. \eqref{eq:marginalised} as the nested sampler explores the astrophysical parameter space.
The direct sampling method is relatively slow compared to importance sampling (by a factor of $\sim250$) requiring the use of \textsc{pBilby} \cite{smith2019expediting}, a parallelized implementation of \textsc{Dynesty} \cite{Dynesty}.

All three methods produce nearly identical posterior distributions, which are difficult to distinguish by eye, illustrating that calibration error distribution has only a very small effect on our inferences about astrophysical parameters. 
This is also verified in Sec. \ref{sec:results} when analysing all events from GWTC-1. 
In Table~\ref{tab:tests}, we present the maximum one-dimensional Jensen–Shannon (JS) divergence \cite{Lin1991} comparing the similarity of the posterior distributions obtained using each method. 
The JS divergence is a symmetric extension of the Kullback-Liebler (KL) divergence \cite{kullback1951} which measures the divergence between 0 bit (no divergence) to 1 bit (maximal divergence). 
We obtain JS divergence values $\lesssim 6\times10^{-3}$ which are similar to those obtained from comparing the results obtained using different stochastic sampling codes to sample the same likelihood~\cite{bilby_gwtc1, bilby, veitch15}.

\begin{table}[t]
    \centering
    \begin{tabular}{|c||c|c|}
        \cline{2-3}
        \multicolumn{1}{c||}{} & direct & importance \\
        \hline\hline
        no error & $\textrm{JS}_{\textrm{RA}} = 8.07\times10^{-3}$ & $\textrm{JS}_{\textrm{RA}} = 5.16\times10^{-3}$ \\\hline
        importance & $\textrm{JS}_{a_1} = 3.94\times10^{-3}$ & \multicolumn{1}{c}{} \\\cline{1-2}
    \end{tabular}
    \caption{
    The largest one-dimensional Jensen–Shannon (JS) divergence (bit) comparing the similarity of the posterior distributions obtained using different methods for a simulated $\text{SNR}=30$ binary black hole signal.
    The small value in each cell indicates that the three methods produce similar distributions, which implies that calibration uncertainty does not have a significant effect on astrophysical inference.
    }
    \label{tab:tests}
\end{table}

The $\textrm{SNR}=200$ event allows us to study what is likely to be the maximum-SNR regime for second-generation gravitational-wave detectors. 
The Advanced LIGO/Virgo network at design sensitivity is expected to observe ${\cal O}(10^4)$ events over its operational lifetime. 
Assuming that the distribution of network SNR scales like $\text{SNR}^{-4}$ \cite{schutz2011networks}, the number of events with a signal-to-noise ratio greater than $200$ will be ${\cal O}(1)$ (see also \cite{chen2014loudest}). 

\begin{figure*}[p]
    \centering
    \includegraphics[width=0.48\linewidth]{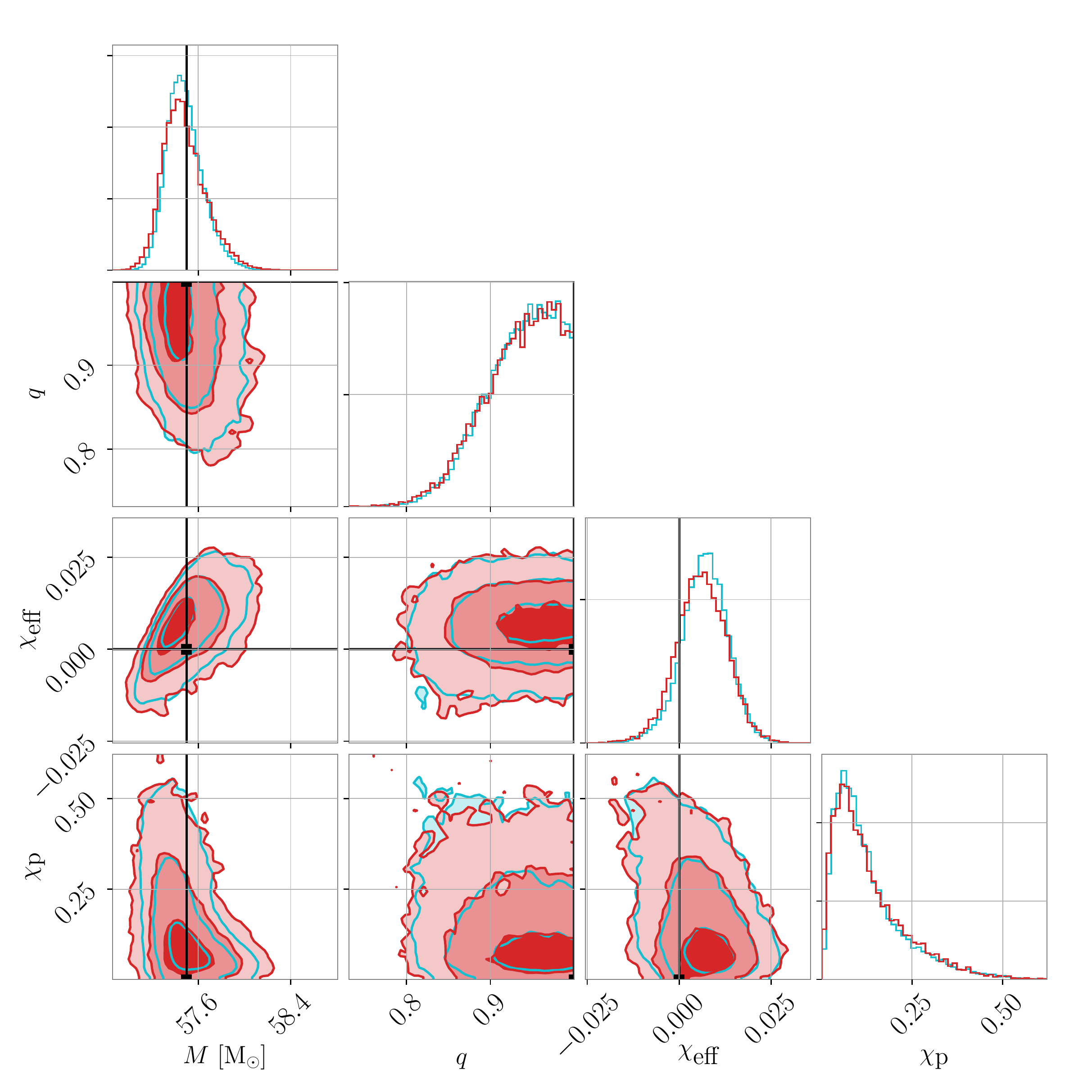}
    \includegraphics[width=0.48\linewidth]{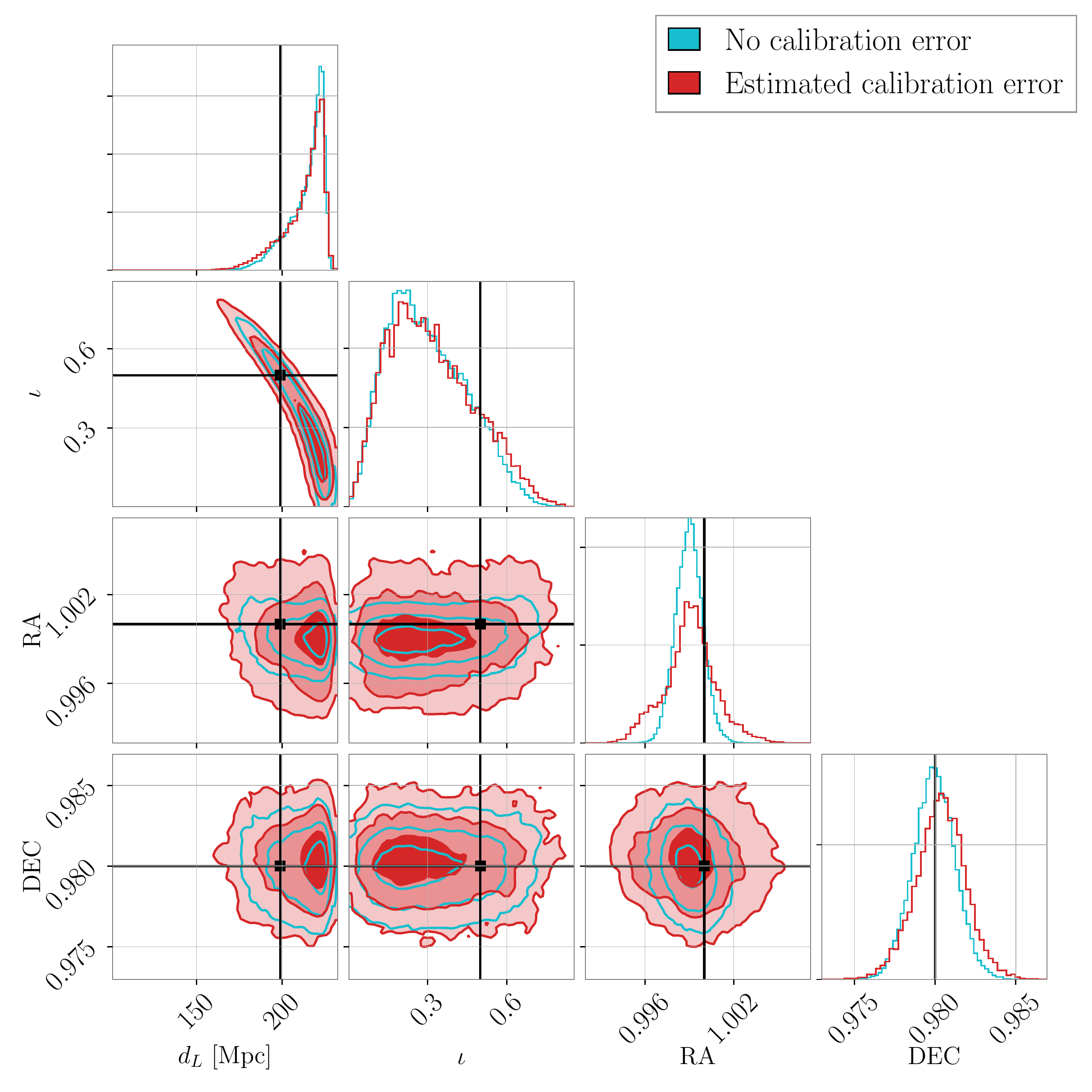}
    \includegraphics[width=\linewidth]{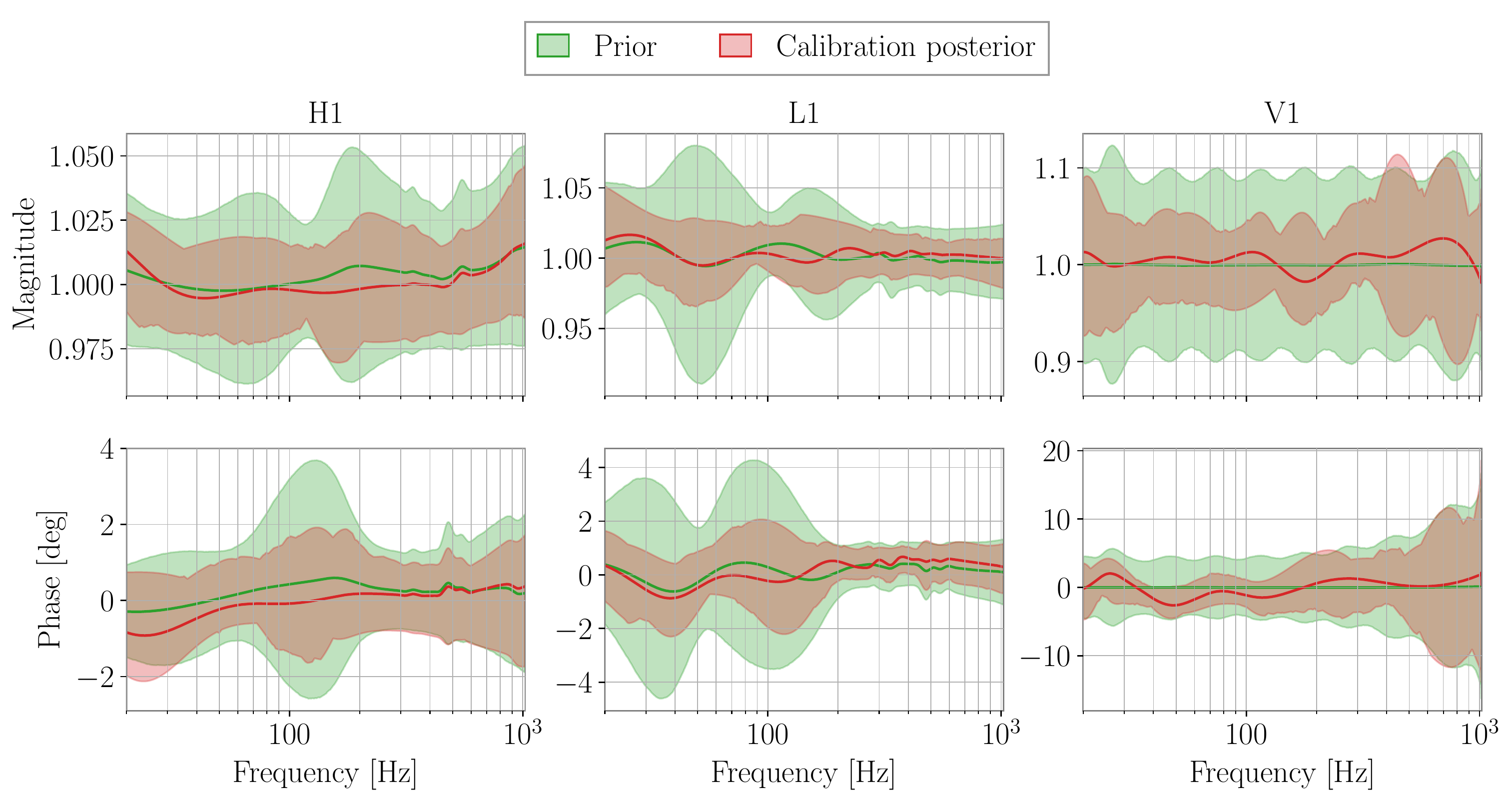}
    \caption{
    Posterior distributions for an SNR$=200$ GW150914-like event.
    The top panels show astrophysical parameters.
    The different shaded regions are $1\sigma, 2\sigma,$ and $3\sigma$ credible intervals.
    The red contours include the calibration error estimate while the blue contours do not.
    The black lines correspond to the injected properties of the source.
    The bottom panels show the reconstructed response function $R$.
    The red curves show the response curves averaged over calibration hyper-parameters.
    The green curves show the response functions averaged over prior samples.
    The 95\% credible intervals are indicated with translucent shading.
    The inclusion of the calibration envelope broadens the majority of astrophysical parameters a modest amount. 
    The sky localization of the event broadens noticeably with the inclusion of calibration uncertainty, expanding by a factor of $\approx2$ in solid angle. 
    This indicates the possibility that even for the loudest events observed, the calibration error estimate may not play a major role in the inferences made about the intrinsic properties of the source.
    It is interesting to note that constraining calibration model parameters at lower frequencies where the gravitational-wave signal is detected can inform the calibration model at higher frequencies where no signal is present.}
    \label{fig:snr_200}
\end{figure*}

The posterior distributions for the astrophysical parameters are presented in the top panels of Fig~\ref{fig:snr_200}.
We see the largest differences in the extrinsic parameters (right panel).
In particular, we highlight that the credible regions on the sky are approximately twice as large when marginalizing over calibration error estimates than when assuming no calibration error is present.
Quantitatively, this difference corresponds to a JS-divergence of 0.105 for right ascension.
More modest changes are seen for the remaining parameters, with JS divergences $\leq 3.04\times10^{-2}$, qualitative astrophysical results are unchanged.
The mean and $95\%$ credible regions for the prior and posterior on the calibration uncertainty are shown in the lower panels.
We note that the width of the posterior credible regions are approximately half that of the prior credible regions.

The Bayes factor for the $\textrm{SNR}=200$ injection is ${\cal B}^\Lambda_{\o} = 2.04\times10^{-4}$, indicating a preference for the null hypothesis that there is no calibration error.
This is expected given that we did not perturb the simulated data to introduce a systematic error. 
However, this result also tells us something interesting about the calibration model.
No calibration error ($\lambda=1$) should be allowed as one possible realisation of the calibration envelope.
What does it mean, therefore, that the data so strongly prefer the null hypothesis for this injection?
We suspect there are two factors at play.
First, some of the preference is likely coming from a large physical calibration model parameter space.
This results in a penalty known as an Occam factor, where simplified models with a smaller prior volume are preferred to models with a larger prior volume, provided the data is fit accurately. 
However, we suspect that there is a more important factor at play: the $10^4$ realizations of the calibration envelope may not be sufficient to adequately fit the zero-error data.
If this is the case, it could be highlighting the limitations that arise when we represent a continuous response function with some finite number of curves.
Additional work beyond our present scope would be useful to investigate these hypotheses. 

\section{Results from GWTC-1}\label{sec:results}
\begin{figure*}[p]
    \centering
    \includegraphics[width=0.48\linewidth]{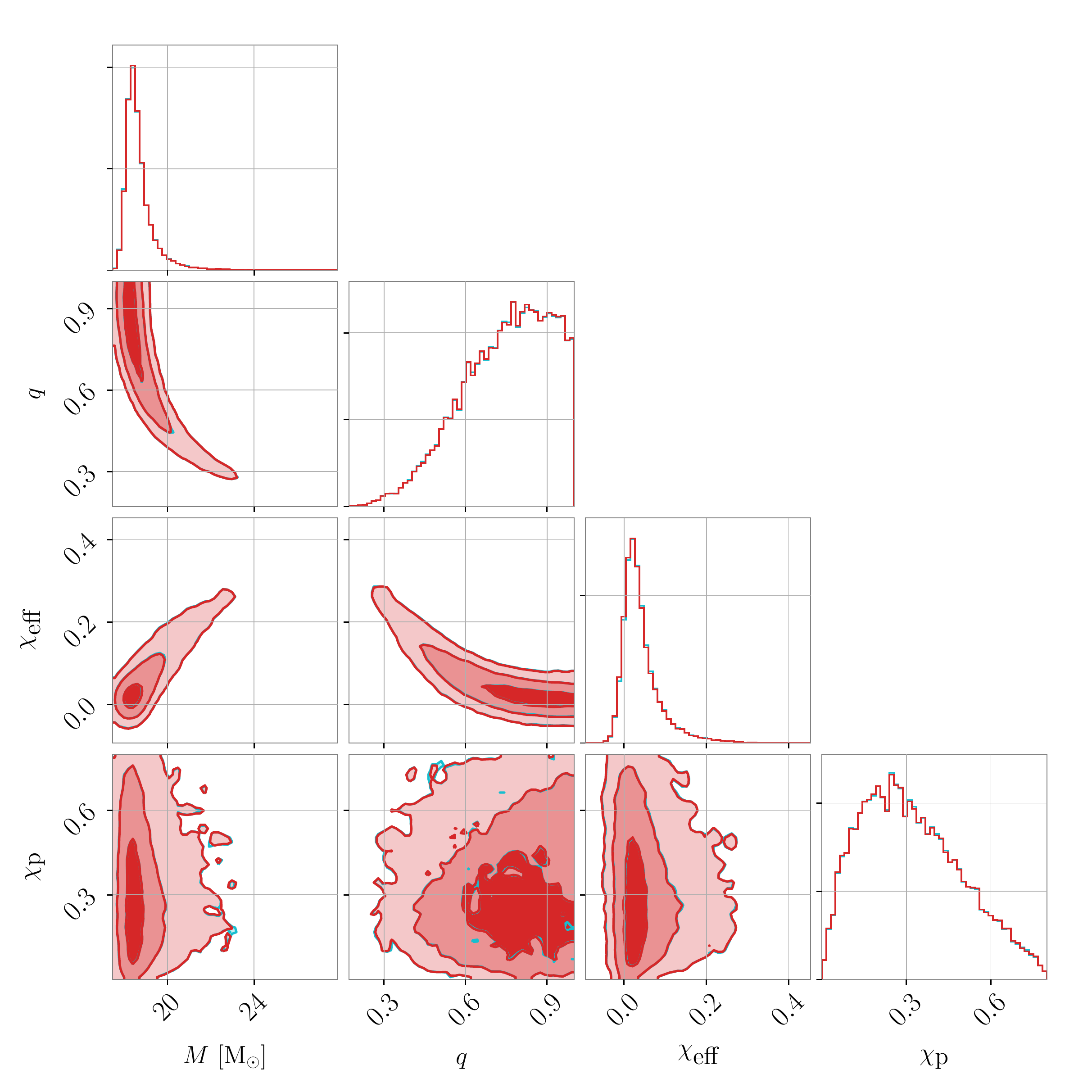}
    \includegraphics[width=0.48\linewidth]{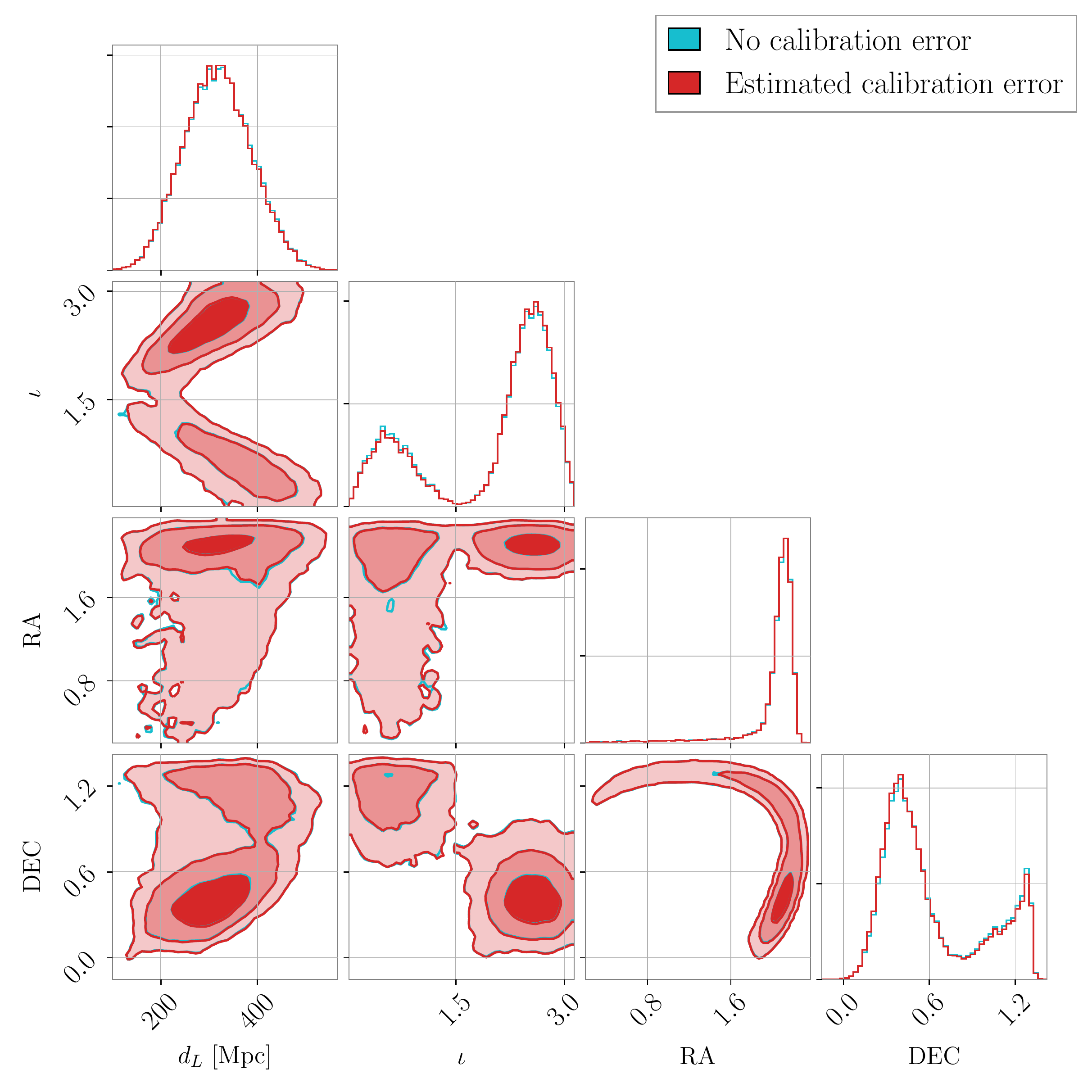}
    \includegraphics[width=\linewidth]{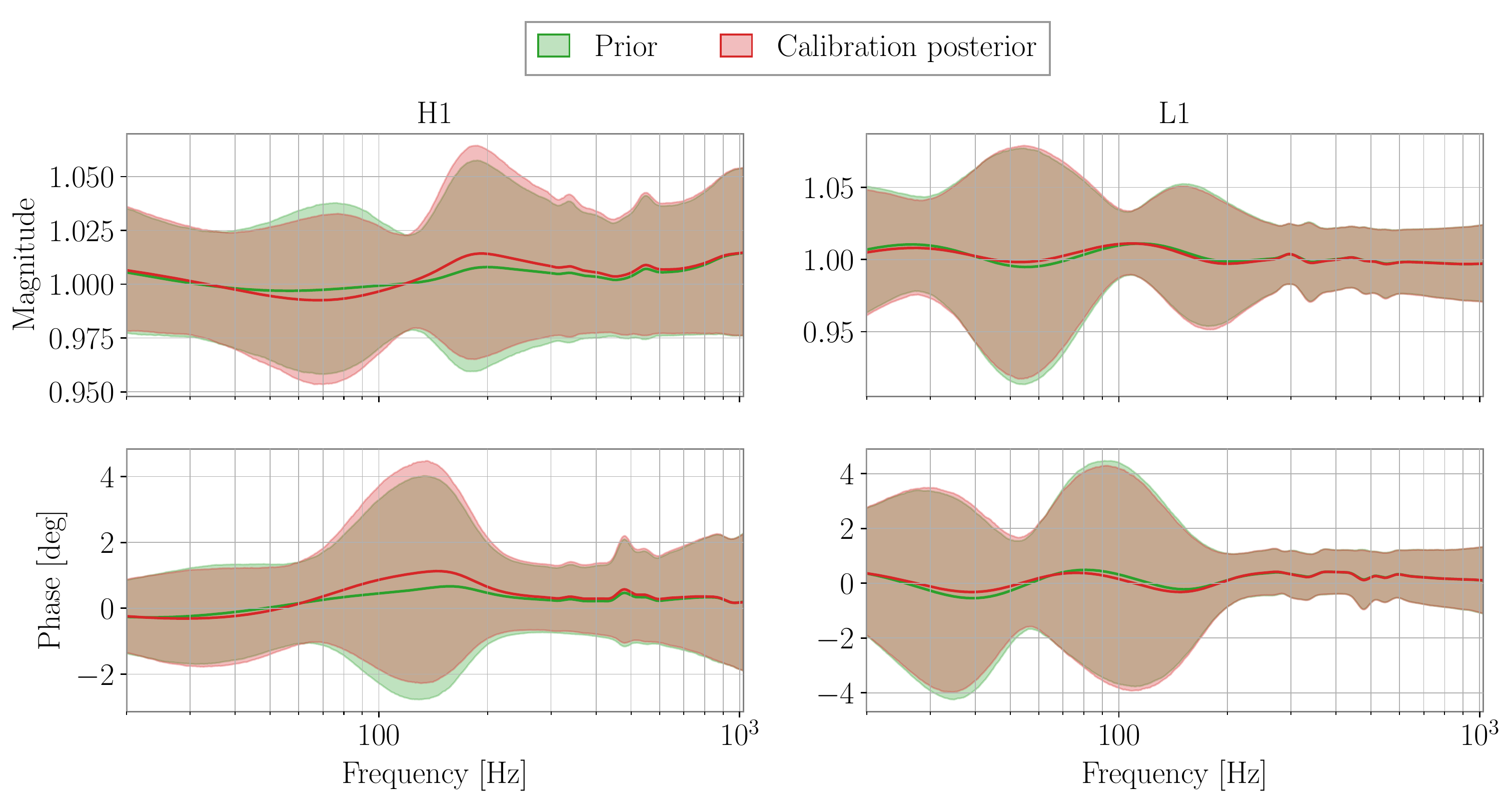}
    \caption{
    Posterior distributions for GW170608.
    The top panels show astrophysical parameters.
    The different shaded regions are $1\sigma, 2\sigma,$ and $3\sigma$ credible intervals.
    The red contours include the calibration error estimate while the blue contours do not.
    The inclusion of uncertainty in the calibration error leads to only marginal changes.
    The bottom panels show the reconstructed response function $R$.
    The red curves show the response curves averaged over calibration hyper-parameters.
    The green curves show the response functions averaged over prior samples.
    The 95\% credible intervals are indicated with translucent shading.
    The data are marginally informative about the calibration parameters.
    }
    \label{fig:GW170608}
\end{figure*}

We analyze the $11$ binary merger events identified in GWTC-1~\cite{gwtc-1, abbott2019open} using the method described in Section~\ref{sec:method}.
Strain data is utilized from the open data release \cite{abbott2019open}, while noise power-spectral densities are used from Ref.~\cite{psds} produced with {\tt \sc BayesWave} \cite{cornish2015bayeswave, littenberg2015bayesian}. 
Calibration error distributions are estimated for LIGO detectors in the first observing run and Virgo using the spline method~\cite{spline_doc, calUncertainties}. Observations during the second observing run directly utilize the physical calibration model presented in Sec. \ref{sec:model}. 
To illustrate the typical effect of the inclusion of the physical calibration model, we first consider GW170608.
In the top panels of Fig.~\ref{fig:GW170608}, we show the posterior distributions for the astrophysical parameters for GW170608.
The red contours include marginalisation over calibration error estimates while the blue contours do not.
While there are small differences between the red and blue contours---we encourage the reader to squint at the posterior distributions for $\text{DEC}$ and $\iota$---it is clear that the inclusion of uncertainty in the calibration error has a very small effect on the size and shape of the astrophysical posterior distributions.
In the bottom panels of Fig.~\ref{fig:GW170608} we show the reconstructed calibration response function.
The thick red curve is averaged over draws from the calibration parameter posterior distribution while the green curve is averaged over draws from the prior.
The slight difference between the red and green credible intervals show a (small) change in the mean and 95\% confidence intervals of the calibration envelope.

We also present the efficiencies and JS divergences for all events in GWTC-1 in Table \ref{tab:results}.
The efficiency for obtaining calibration-marginalized samples is $\epsilon=78.2\%$ for GW150914, and $\epsilon=64.9\%$ for GW170817.
The non-unity efficiency for these two events is due to their larger network SNR.
For other events in GWTC-1, we obtain efficiencies of $\epsilon={97.4-99.7}\%$.
Visual inspection of the posterior distributions for the other events in GWTC-1 confirm that the effect of uncertainty in the calibration error is negligible for events in GWTC-1.
This is further verified by JS divergences $\lesssim 1.5\times10^{-3}$, which are comparable to values found between different implementations of stochastic sampling algorithms \cite{bilby_gwtc1} and smaller than differences due to differences in waveform models~\cite{gwtc-1}.
The full analysis of GWTC-1 results are available for download~\cite{online}.

\begin{table}
    \centering
    \begin{tabular}{|c|c|c|c|}
        \hline
        event & $\epsilon$ (\%) & ${\cal B}^\Lambda_{\o}$ & Max. JS divergence (bit) \\
        \hline\hline
        GW150914 & 78.2 & 0.97 & $\textrm{JS}_{t_c} = 1.55\times10^{-3}$\\\hline
        GW151012 & 99.7 & 0.97 & $\textrm{JS}_{\mathcal{M}} = 5.05\times10^{-5}$\\\hline
        GW151226 & 99.4 & 0.96 & $\textrm{JS}_{\mathcal{M}} = 1.71\times10^{-4}$\\\hline
        GW170104 & 98.7 & 0.96 & $\textrm{JS}_{\phi_{JL}} = 2.87\times10^{-5}$\\\hline
        GW170608 & 97.4 & 1.12 & $\textrm{JS}_{\textrm{DEC}} = 2.98\times10^{-4}$\\\hline
        GW170729 & 99.2 & 0.93 & $\textrm{JS}_{\mathcal{M}} = 1.12\times10^{-4}$\\\hline
        GW170809 & 99.3 & 0.91 & $\textrm{JS}_{t_c} = 1.28\times10^{-4}$\\\hline
        GW170814 & 98.0 & 1.08 & $\textrm{JS}_{\textrm{RA}} = 2.93\times10^{-4}$\\\hline
        GW170817 & 64.9 & 1.93 & $\textrm{JS}_{d_L} = 8.90\times10^{-4}$\\\hline
        GW170818 & 98.9 & 1.06 & $\textrm{JS}_{\phi_{JL}} = 2.88\times10^{-4}$\\\hline
        GW170823 & 98.9 & 0.97 & $\textrm{JS}_{\textrm{RA}} = 2.89\times10^{-5}$\\\hline
        \hline
        $\text{SNR}=30$ inj & 90.3 & 0.78 & $\textrm{JS}_{\textrm{RA}} = 5.16\times10^{-3}$ \\\hline
        $\text{SNR}=200$ inj & -- & $2.04\times10^{-4}$ & $\textrm{JS}_{\textrm{RA}} = 1.05\times10^{-1}$ \\\hline
    \end{tabular}
    \caption{
    Summary of results from GWTC-1 and the two injections described in Section~\ref{sec:testing}.
    The efficiency, $\epsilon$, is defined in Eq.~\ref{eq:efficiency}.
    The Bayes factor, ${\cal B}^\Lambda_{\o}$, compares the likelihood obtained marginalizing over the calibration envelope to the marginal likelihood obtained ignoring any calibration error estimates.
    The Jensen-Shannon (JS) divergence measures the change in the posterior distribution when we include calibration error estimate.
    For the $\textrm{SNR}=200$ injection, no efficiency is given as the results were obtained via direct sampling of the marginalized likelihood.
    }
    \label{tab:results}
\end{table}

\begin{figure*}[t!]
    \centering
    \includegraphics[width=\linewidth]{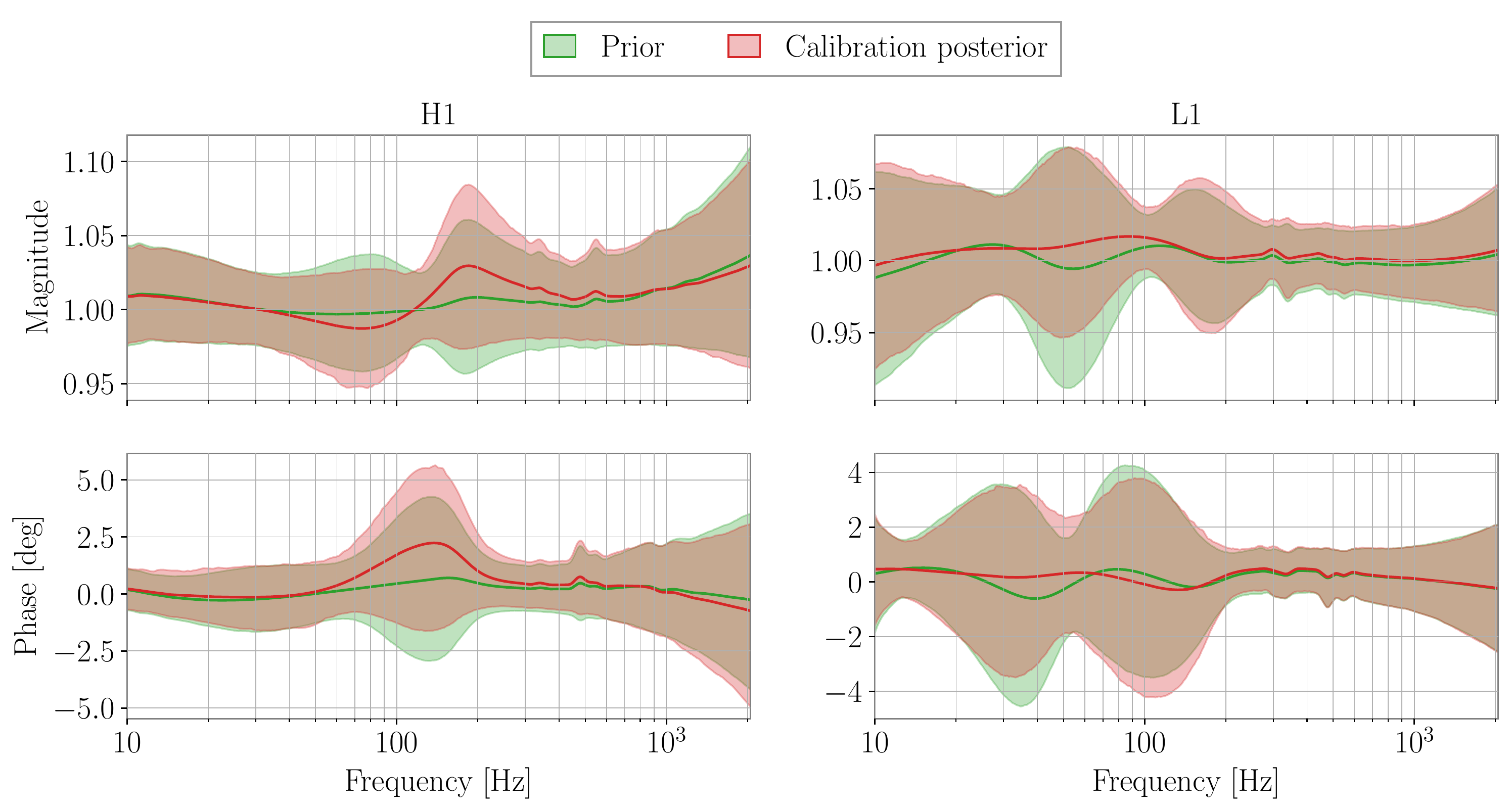}
    \caption{Calibration response curve posterior distributions for both LIGO detectors informed by all events during the second observing run evaluated at the time of GW170729. 
    Marginal shifts in the calibration error distributions are observed. 
    The Bayes factor marginally favors the inclusion of the calibration error distribution over the zero-error hypothesis by $2.33$. 
    }
    \label{fig:stacked_calibration}
\end{figure*}

Finally, we conclude by determining the calibration envelope using events from the second observing run (O2) as standard sirens. 
We only use events from O2 to ensure that the time-independent calibration parameters are identical. 
We compute the combined weights for the calibration response curves following Eq.~\eqref{eq:stacking}.
With eight events in O2, the combined calibration envelope is only marginally informed by the gravitational-wave signals. 
The reconstructed envelope, evaluated at the time of GW170729, is presented in Fig.~\ref{fig:stacked_calibration}. 
We observe only a modest change from the prior. 
The total Bayes factor comparing the calibration uncertainty hypothesis to the zero-uncertainty hypothesis is likewise modest: ${\cal B}_{\o}^\Lambda=2.33$.
More events are required to meaningfully inform the calibration error estimate.
However, with the requirement to periodically update the calibration model parameters as improvements to the detectors are made, the required number of events may not be achievable in the foreseeable future. 
This is also concluded within Ref. \cite{Essick2019}, where they comment that due to the periodic model updates, astrophysical calibration may never be competitive. 

\section{Conclusions}\label{sec:conclusions}
We have presented a calibration-marginalized likelihood for astrophysical parameters employing a physically informed model for the calibration error as presented in~\cite{Cahillane2017,Sun2020}. 
Within the signal-to-noise ratio regime of previously observed events and estimates of calibration errors at the levels reported in GWTC-1, we find the effect of calibration error is at the same level as the effect of stochastic sampling errors and less than other known systematics. 
Recent work from Ref.~\cite{MIT} has also investigated similar marginalization using direct sampling of the calibration error curve index, instead of importance sampling.
The conclusions drawn within  Ref.~\cite{MIT} are consistent with those drawn here. 
We also demonstrated that, if calibration errors remain as low as in GWTC-1, even future loud events will incur only modest changes in the estimates of astrophysical parameters, with the potential exception of increased uncertainty in the sky location.
We also demonstrated the improved inference of calibration parameters using the collection of events from GWTC-1 as standard sirens.
Our findings are consistent with~\cite{Essick2019}, where it is found that using gravitational-wave events to improve the estimate of calibration errors beyond that determined from in-situ measurements requires thousands of detections. 

\section*{Acknowledgements}
We thank Salvatore Vitale, Carl-Johan Haster, Lilli Sun, Ben Farr, and Evan Goetz for insightful comments and sharing an early version of their manuscript.
We thank Nikhil Sarin and Rory Smith in providing guidance for the use of \textsc{pBilby}.
We thank Greg Mendell and Rick Savage for thoughtful discussions, and Reed Essick for helpful comments on the manuscript.
This work is supported through Australian Research Council (ARC) Centre of Excellence CE170100004. 
EP acknowledges the support of the LSC Fellows program. 
PDL is supported through ARC Future Fellowship FT160100112, and ARC Discovery Project DP180103155.
ET is supported through ARC Future Fellowship FT150100281.
This is document LIGO-P2000294.

This research has made use of data, software and/or web tools obtained from the Gravitational Wave Open Science Center (https://www.gw-openscience.org), a service of LIGO Laboratory, the LIGO Scientific Collaboration and the Virgo Collaboration.The authors are grateful for computational resources provided by the LIGO Laboratory and supported by National Science Foundation Grants PHY-- 0757058 and PHY-- 0823459. 
Computing was performed computing clusters at California Institute of Technology (LIGO Laboratory) and Swinburne University of Technology (OzSTAR).
We would like to thank all of the essential workers who put their health at risk during the COVID-19 pandemic, without whom we would not have been able to complete this work.

\bibliography{refs}

\begin{thebibliography}{58}
\expandafter\ifx\csname natexlab\endcsname\relax\def\natexlab#1{#1}\fi
\expandafter\ifx\csname bibnamefont\endcsname\relax
  \def\bibnamefont#1{#1}\fi
\expandafter\ifx\csname bibfnamefont\endcsname\relax
  \def\bibfnamefont#1{#1}\fi
\expandafter\ifx\csname citenamefont\endcsname\relax
  \def\citenamefont#1{#1}\fi
\expandafter\ifx\csname url\endcsname\relax
  \def\url#1{\texttt{#1}}\fi
\expandafter\ifx\csname urlprefix\endcsname\relax\def\urlprefix{URL }\fi
\providecommand{\bibinfo}[2]{#2}
\providecommand{\eprint}[2][]{\url{#2}}

\bibitem[{\citenamefont{Abbott et~al.}(2016)}]{GW150914}
\bibinfo{author}{\bibfnamefont{B.~P.} \bibnamefont{Abbott}}
  \bibnamefont{et~al.} (\bibinfo{collaboration}{LIGO Scientific and Virgo
  Collaborations}), \bibinfo{journal}{Phys. Rev. Lett.}
  \textbf{\bibinfo{volume}{116}}, \bibinfo{pages}{061102}
  (\bibinfo{year}{2016}).

\bibitem[{\citenamefont{Abbott et~al.}(2017{\natexlab{a}})}]{GW170817}
\bibinfo{author}{\bibfnamefont{B.~P.} \bibnamefont{Abbott}}
  \bibnamefont{et~al.} (\bibinfo{collaboration}{LIGO Scientific and Virgo
  Collaborations}), \bibinfo{journal}{Phys. Rev. Lett.}
  \textbf{\bibinfo{volume}{119}}, \bibinfo{pages}{161101}
  (\bibinfo{year}{2017}{\natexlab{a}}).

\bibitem[{\citenamefont{Abbott et~al.}(2019{\natexlab{a}})}]{gwtc-1}
\bibinfo{author}{\bibfnamefont{B.~P.} \bibnamefont{Abbott}}
  \bibnamefont{et~al.} (\bibinfo{collaboration}{LIGO Scientific and Virgo
  Collaborations}), \bibinfo{journal}{Phys. Rev. X}
  \textbf{\bibinfo{volume}{9}}, \bibinfo{pages}{031040}
  (\bibinfo{year}{2019}{\natexlab{a}}).

\bibitem[{\citenamefont{Abbott et~al.}(2017{\natexlab{b}})}]{hubbleGW}
\bibinfo{author}{\bibfnamefont{B.~P.} \bibnamefont{Abbott}}
  \bibnamefont{et~al.} (\bibinfo{collaboration}{LIGO Scientific and Virgo
  Collaborations}), \bibinfo{journal}{Nature} \textbf{\bibinfo{volume}{551}},
  \bibinfo{pages}{85} (\bibinfo{year}{2017}{\natexlab{b}}).

\bibitem[{\citenamefont{Abbott et~al.}(2019{\natexlab{b}})}]{gwtc1pop}
\bibinfo{author}{\bibfnamefont{B.~P.} \bibnamefont{Abbott}}
  \bibnamefont{et~al.} (\bibinfo{collaboration}{LIGO Scientific and Virgo
  Collaborations}), \bibinfo{journal}{The Astrophysical Journal Letters}
  \textbf{\bibinfo{volume}{882}}, \bibinfo{pages}{L24}
  (\bibinfo{year}{2019}{\natexlab{b}}).

\bibitem[{\citenamefont{Abbott et~al.}(2019{\natexlab{c}})}]{gwtc1GR}
\bibinfo{author}{\bibfnamefont{B.~P.} \bibnamefont{Abbott}}
  \bibnamefont{et~al.} (\bibinfo{collaboration}{LIGO Scientific and Virgo
  Collaborations}), \bibinfo{journal}{Physical Review D}
  \textbf{\bibinfo{volume}{100}}, \bibinfo{pages}{104036}
  (\bibinfo{year}{2019}{\natexlab{c}}).

\bibitem[{\citenamefont{Schutz}(2011)}]{schutz2011networks}
\bibinfo{author}{\bibfnamefont{B.~F.} \bibnamefont{Schutz}},
  \bibinfo{journal}{Classical and Quantum Gravity}
  \textbf{\bibinfo{volume}{28}}, \bibinfo{pages}{125023}
  (\bibinfo{year}{2011}).

\bibitem[{\citenamefont{Vitale}(2016)}]{vitale2016}
\bibinfo{author}{\bibfnamefont{S.}~\bibnamefont{Vitale}},
  \bibinfo{journal}{Phys. Rev. D} \textbf{\bibinfo{volume}{94}},
  \bibinfo{pages}{121501} (\bibinfo{year}{2016}).

\bibitem[{\citenamefont{Abbott
  et~al.}(2019{\natexlab{d}})}]{ligo2019gravitational}
\bibinfo{author}{\bibfnamefont{B.~P.} \bibnamefont{Abbott}}
  \bibnamefont{et~al.} (\bibinfo{collaboration}{LIGO Scientific and Virgo
  Collaborations}), \bibinfo{journal}{arxiv/1904.03187}
  (\bibinfo{year}{2019}{\natexlab{d}}).

\bibitem[{\citenamefont{P{\"u}rrer and Haster}(2020)}]{Purrer2020}
\bibinfo{author}{\bibfnamefont{M.}~\bibnamefont{P{\"u}rrer}} \bibnamefont{and}
  \bibinfo{author}{\bibfnamefont{C.-J.} \bibnamefont{Haster}},
  \bibinfo{journal}{Phys. Rev. Res.} \textbf{\bibinfo{volume}{2}},
  \bibinfo{pages}{023151} (\bibinfo{year}{2020}).

\bibitem[{\citenamefont{Biscoveanu et~al.}(2020)\citenamefont{Biscoveanu,
  Haster, Vitale, and Davies}}]{sylviaPSD}
\bibinfo{author}{\bibfnamefont{S.}~\bibnamefont{Biscoveanu}},
  \bibinfo{author}{\bibfnamefont{C.-J.} \bibnamefont{Haster}},
  \bibinfo{author}{\bibfnamefont{S.}~\bibnamefont{Vitale}}, \bibnamefont{and}
  \bibinfo{author}{\bibfnamefont{J.}~\bibnamefont{Davies}},
  \bibinfo{journal}{Phys. Rev. D} \textbf{\bibinfo{volume}{102}},
  \bibinfo{pages}{023008} (\bibinfo{year}{2020}).

\bibitem[{\citenamefont{Talbot and Thrane}(2020)}]{student-t}
\bibinfo{author}{\bibfnamefont{C.}~\bibnamefont{Talbot}} \bibnamefont{and}
  \bibinfo{author}{\bibfnamefont{E.}~\bibnamefont{Thrane}}
  (\bibinfo{year}{2020}), \bibinfo{note}{arxiv/2006.05292}.

\bibitem[{\citenamefont{Chatziioannou et~al.}(2019)\citenamefont{Chatziioannou,
  Haster, Littenberg, Farr, Ghonge, Millhouse, Clark, and
  Cornish}}]{chatziioannou2019noise}
\bibinfo{author}{\bibfnamefont{K.}~\bibnamefont{Chatziioannou}},
  \bibinfo{author}{\bibfnamefont{C.-J.} \bibnamefont{Haster}},
  \bibinfo{author}{\bibfnamefont{T.~B.} \bibnamefont{Littenberg}},
  \bibinfo{author}{\bibfnamefont{W.~M.} \bibnamefont{Farr}},
  \bibinfo{author}{\bibfnamefont{S.}~\bibnamefont{Ghonge}},
  \bibinfo{author}{\bibfnamefont{M.}~\bibnamefont{Millhouse}},
  \bibinfo{author}{\bibfnamefont{J.~A.} \bibnamefont{Clark}}, \bibnamefont{and}
  \bibinfo{author}{\bibfnamefont{N.}~\bibnamefont{Cornish}},
  \bibinfo{journal}{Physical Review D} \textbf{\bibinfo{volume}{100}},
  \bibinfo{pages}{104004} (\bibinfo{year}{2019}).

\bibitem[{\citenamefont{Vitale et~al.}(2012)\citenamefont{Vitale, Pozzo, Li,
  Broeck, Aylott, and Veitch}}]{Vitale2014}
\bibinfo{author}{\bibfnamefont{S.}~\bibnamefont{Vitale}},
  \bibinfo{author}{\bibfnamefont{W.~D.} \bibnamefont{Pozzo}},
  \bibinfo{author}{\bibfnamefont{T.~G.~F.} \bibnamefont{Li}},
  \bibinfo{author}{\bibfnamefont{C.~V.~D.} \bibnamefont{Broeck}},
  \bibinfo{author}{\bibfnamefont{B.}~\bibnamefont{Aylott}}, \bibnamefont{and}
  \bibinfo{author}{\bibfnamefont{J.}~\bibnamefont{Veitch}},
  \bibinfo{journal}{Phys. Rev. D} \textbf{\bibinfo{volume}{85}},
  \bibinfo{pages}{064034} (\bibinfo{year}{2012}).

\bibitem[{\citenamefont{Abbott
  et~al.}(2017{\natexlab{c}})}]{abbott2017calibration}
\bibinfo{author}{\bibfnamefont{B.~P.} \bibnamefont{Abbott}}
  \bibnamefont{et~al.} (\bibinfo{collaboration}{LIGO Scientific and Virgo
  Collaborations}), \bibinfo{journal}{Physical Review D}
  \textbf{\bibinfo{volume}{95}}, \bibinfo{pages}{062003}
  (\bibinfo{year}{2017}{\natexlab{c}}).

\bibitem[{\citenamefont{Cahillane et~al.}(2017)}]{Cahillane2017}
\bibinfo{author}{\bibfnamefont{C.}~\bibnamefont{Cahillane}}
  \bibnamefont{et~al.}, \bibinfo{journal}{Phys. Rev. D}
  \textbf{\bibinfo{volume}{96}}, \bibinfo{pages}{102001}
  (\bibinfo{year}{2017}).

\bibitem[{\citenamefont{Acernese et~al.}(2018)}]{acernese2018calibration}
\bibinfo{author}{\bibfnamefont{F.}~\bibnamefont{Acernese}} \bibnamefont{et~al.}
  (\bibinfo{collaboration}{Virgo Collaboration}), \bibinfo{journal}{Classical
  and Quantum Gravity} \textbf{\bibinfo{volume}{35}}, \bibinfo{pages}{205004}
  (\bibinfo{year}{2018}).

\bibitem[{\citenamefont{Sun et~al.}(2020)}]{Sun2020}
\bibinfo{author}{\bibfnamefont{L.}~\bibnamefont{Sun}} \bibnamefont{et~al.},
  \bibinfo{journal}{Class. Quant. Grav.}  (\bibinfo{year}{2020}).

\bibitem[{\citenamefont{{Farr} et~al.}(2014)\citenamefont{{Farr}, Farr, and
  Littenberg}}]{spline_doc}
\bibinfo{author}{\bibfnamefont{W.~M.} \bibnamefont{{Farr}}},
  \bibinfo{author}{\bibfnamefont{B.}~\bibnamefont{Farr}}, \bibnamefont{and}
  \bibinfo{author}{\bibfnamefont{T.}~\bibnamefont{Littenberg}},
  \bibinfo{type}{Tech. Rep.} \bibinfo{number}{LIGO-T1400682}
  (\bibinfo{year}{2014}),
  \urlprefix\url{https://dcc.ligo.org/LIGO-T1400682/public}.

\bibitem[{\citenamefont{Payne et~al.}(2019)\citenamefont{Payne, Talbot, and
  Thrane}}]{hom}
\bibinfo{author}{\bibfnamefont{E.}~\bibnamefont{Payne}},
  \bibinfo{author}{\bibfnamefont{C.}~\bibnamefont{Talbot}}, \bibnamefont{and}
  \bibinfo{author}{\bibfnamefont{E.}~\bibnamefont{Thrane}},
  \bibinfo{journal}{Phys. Rev. D} \textbf{\bibinfo{volume}{100}},
  \bibinfo{pages}{123017} (\bibinfo{year}{2019}).

\bibitem[{\citenamefont{Robert and Casella}(2004)}]{Robert&Casella}
\bibinfo{author}{\bibfnamefont{C.~P.} \bibnamefont{Robert}} \bibnamefont{and}
  \bibinfo{author}{\bibfnamefont{G.}~\bibnamefont{Casella}},
  \emph{\bibinfo{title}{Monte Carlo Statistical Methods}}
  (\bibinfo{publisher}{Springer}, \bibinfo{address}{New York},
  \bibinfo{year}{2004}), \bibinfo{edition}{2nd} ed.

\bibitem[{\citenamefont{Liu}(2004)}]{Liu}
\bibinfo{author}{\bibfnamefont{J.~S.} \bibnamefont{Liu}},
  \emph{\bibinfo{title}{Monte Carlo Strategies in Scientific Computing}}
  (\bibinfo{publisher}{Springer}, \bibinfo{address}{New York},
  \bibinfo{year}{2004}), \bibinfo{edition}{1st} ed.

\bibitem[{\citenamefont{Essick and Holz}(2019)}]{Essick2019}
\bibinfo{author}{\bibfnamefont{R.}~\bibnamefont{Essick}} \bibnamefont{and}
  \bibinfo{author}{\bibfnamefont{D.~E.} \bibnamefont{Holz}},
  \bibinfo{journal}{Class. Quant. Grav.} \textbf{\bibinfo{volume}{36}},
  \bibinfo{pages}{125002} (\bibinfo{year}{2019}).

\bibitem[{\citenamefont{{Aasi} et~al.}(2015)}]{aLIGO}
\bibinfo{author}{\bibfnamefont{J.}~\bibnamefont{{Aasi}}} \bibnamefont{et~al.}
  (\bibinfo{collaboration}{LIGO Collaboration}), \bibinfo{journal}{Class.
  Quant. Grav.} \textbf{\bibinfo{volume}{32}}, \bibinfo{pages}{074001}
  (\bibinfo{year}{2015}).

\bibitem[{\citenamefont{Chen and Holz}(2014)}]{chen2014loudest}
\bibinfo{author}{\bibfnamefont{H.-Y.} \bibnamefont{Chen}} \bibnamefont{and}
  \bibinfo{author}{\bibfnamefont{D.~E.} \bibnamefont{Holz}},
  \bibinfo{journal}{arxiv/1409.0522}  (\bibinfo{year}{2014}).

\bibitem[{\citenamefont{Acernese et~al.}(2014)}]{acernese2014advanced}
\bibinfo{author}{\bibfnamefont{F.}~\bibnamefont{Acernese}} \bibnamefont{et~al.}
  (\bibinfo{collaboration}{Virgo Collaboration}), \bibinfo{journal}{Classical
  and Quantum Gravity} \textbf{\bibinfo{volume}{32}}, \bibinfo{pages}{024001}
  (\bibinfo{year}{2014}).

\bibitem[{\citenamefont{Izumi and Sigg}(2016)}]{Izumi2016}
\bibinfo{author}{\bibfnamefont{K.}~\bibnamefont{Izumi}} \bibnamefont{and}
  \bibinfo{author}{\bibfnamefont{D.}~\bibnamefont{Sigg}},
  \bibinfo{journal}{Classical and Quantum Gravity}
  \textbf{\bibinfo{volume}{34}}, \bibinfo{pages}{015001}
  (\bibinfo{year}{2016}).

\bibitem[{\citenamefont{Robertson et~al.}(2002)}]{robertson2002quadruple}
\bibinfo{author}{\bibfnamefont{N.}~\bibnamefont{Robertson}}
  \bibnamefont{et~al.}, \bibinfo{journal}{Classical and Quantum Gravity}
  \textbf{\bibinfo{volume}{19}}, \bibinfo{pages}{4043} (\bibinfo{year}{2002}).

\bibitem[{\citenamefont{Aston et~al.}(2012)}]{aston2012update}
\bibinfo{author}{\bibfnamefont{S.}~\bibnamefont{Aston}} \bibnamefont{et~al.},
  \bibinfo{journal}{Classical and Quantum Gravity}
  \textbf{\bibinfo{volume}{29}}, \bibinfo{pages}{235004}
  (\bibinfo{year}{2012}).

\bibitem[{\citenamefont{Carbone et~al.}(2012)}]{carbone2012sensors}
\bibinfo{author}{\bibfnamefont{L.}~\bibnamefont{Carbone}} \bibnamefont{et~al.},
  \bibinfo{journal}{Classical and Quantum Gravity}
  \textbf{\bibinfo{volume}{29}}, \bibinfo{pages}{115005}
  (\bibinfo{year}{2012}).

\bibitem[{\citenamefont{Miyakawa et~al.}(2006)}]{miyakawa2006measurement}
\bibinfo{author}{\bibfnamefont{O.}~\bibnamefont{Miyakawa}}
  \bibnamefont{et~al.}, \bibinfo{journal}{Physical Review D}
  \textbf{\bibinfo{volume}{74}}, \bibinfo{pages}{022001}
  (\bibinfo{year}{2006}).

\bibitem[{\citenamefont{Foreman-Mackey
  et~al.}(2013)\citenamefont{Foreman-Mackey, Hogg, Lang, and Goodman}}]{emcee}
\bibinfo{author}{\bibfnamefont{D.}~\bibnamefont{Foreman-Mackey}},
  \bibinfo{author}{\bibfnamefont{D.~W.} \bibnamefont{Hogg}},
  \bibinfo{author}{\bibfnamefont{D.}~\bibnamefont{Lang}}, \bibnamefont{and}
  \bibinfo{author}{\bibfnamefont{J.}~\bibnamefont{Goodman}},
  \bibinfo{journal}{Publications of the Astronomical Society of the Pacific}
  \textbf{\bibinfo{volume}{125}}, \bibinfo{pages}{306} (\bibinfo{year}{2013}).

\bibitem[{\citenamefont{Karki et~al.}(2016)}]{karki2016advanced}
\bibinfo{author}{\bibfnamefont{S.}~\bibnamefont{Karki}} \bibnamefont{et~al.},
  \bibinfo{journal}{Review of Scientific Instruments}
  \textbf{\bibinfo{volume}{87}}, \bibinfo{pages}{114503}
  (\bibinfo{year}{2016}).

\bibitem[{\citenamefont{Bhattacharjee et~al.}(2020)\citenamefont{Bhattacharjee,
  Lecoeuche, Karki, Betzwieser, Bossilkov, Kandhasamy, Payne, and
  Savage}}]{bhattacharjee2020fiducial}
\bibinfo{author}{\bibfnamefont{D.}~\bibnamefont{Bhattacharjee}},
  \bibinfo{author}{\bibfnamefont{Y.}~\bibnamefont{Lecoeuche}},
  \bibinfo{author}{\bibfnamefont{S.}~\bibnamefont{Karki}},
  \bibinfo{author}{\bibfnamefont{J.}~\bibnamefont{Betzwieser}},
  \bibinfo{author}{\bibfnamefont{V.}~\bibnamefont{Bossilkov}},
  \bibinfo{author}{\bibfnamefont{S.}~\bibnamefont{Kandhasamy}},
  \bibinfo{author}{\bibfnamefont{E.}~\bibnamefont{Payne}}, \bibnamefont{and}
  \bibinfo{author}{\bibfnamefont{R.}~\bibnamefont{Savage}},
  \bibinfo{journal}{arxiv/2006.00130}  (\bibinfo{year}{2020}).

\bibitem[{\citenamefont{Seeger}(2004)}]{gpr}
\bibinfo{author}{\bibfnamefont{M.}~\bibnamefont{Seeger}},
  \bibinfo{journal}{International journal of neural systems}
  \textbf{\bibinfo{volume}{14}}, \bibinfo{pages}{69} (\bibinfo{year}{2004}).

\bibitem[{\citenamefont{Pedregosa et~al.}(2011)}]{scikit_gpr}
\bibinfo{author}{\bibfnamefont{F.}~\bibnamefont{Pedregosa}}
  \bibnamefont{et~al.}, \bibinfo{journal}{the Journal of machine Learning
  research} \textbf{\bibinfo{volume}{12}}, \bibinfo{pages}{2825}
  (\bibinfo{year}{2011}).

\bibitem[{\citenamefont{Thrane and Talbot}(2019)}]{intro}
\bibinfo{author}{\bibfnamefont{E.}~\bibnamefont{Thrane}} \bibnamefont{and}
  \bibinfo{author}{\bibfnamefont{C.}~\bibnamefont{Talbot}},
  \bibinfo{journal}{Pub. Astron. Soc. Aust.} \textbf{\bibinfo{volume}{36}},
  \bibinfo{pages}{E010} (\bibinfo{year}{2019}).

\bibitem[{\citenamefont{Husa et~al.}(2016)\citenamefont{Husa, Khan, Hannam,
  P\"urrer, Ohme, Forteza, and Boh\'e}}]{imrphenompv2_1}
\bibinfo{author}{\bibfnamefont{S.}~\bibnamefont{Husa}},
  \bibinfo{author}{\bibfnamefont{S.}~\bibnamefont{Khan}},
  \bibinfo{author}{\bibfnamefont{M.}~\bibnamefont{Hannam}},
  \bibinfo{author}{\bibfnamefont{M.}~\bibnamefont{P\"urrer}},
  \bibinfo{author}{\bibfnamefont{F.}~\bibnamefont{Ohme}},
  \bibinfo{author}{\bibfnamefont{X.~J.} \bibnamefont{Forteza}},
  \bibnamefont{and} \bibinfo{author}{\bibfnamefont{A.}~\bibnamefont{Boh\'e}},
  \bibinfo{journal}{Phys. Rev. D} \textbf{\bibinfo{volume}{93}},
  \bibinfo{pages}{044006} (\bibinfo{year}{2016}).

\bibitem[{\citenamefont{Khan et~al.}(2016)\citenamefont{Khan, Husa, Hannam,
  Ohme, P\"urrer, Forteza, and Boh\'e}}]{imrphenompv2_2}
\bibinfo{author}{\bibfnamefont{S.}~\bibnamefont{Khan}},
  \bibinfo{author}{\bibfnamefont{S.}~\bibnamefont{Husa}},
  \bibinfo{author}{\bibfnamefont{M.}~\bibnamefont{Hannam}},
  \bibinfo{author}{\bibfnamefont{F.}~\bibnamefont{Ohme}},
  \bibinfo{author}{\bibfnamefont{M.}~\bibnamefont{P\"urrer}},
  \bibinfo{author}{\bibfnamefont{X.~J.} \bibnamefont{Forteza}},
  \bibnamefont{and} \bibinfo{author}{\bibfnamefont{A.}~\bibnamefont{Boh\'e}},
  \bibinfo{journal}{Phys. Rev. D} \textbf{\bibinfo{volume}{93}},
  \bibinfo{pages}{044007} (\bibinfo{year}{2016}).

\bibitem[{\citenamefont{Dietrich et~al.}(2017)\citenamefont{Dietrich, Bernuzzi,
  and Tichy}}]{dietrich2017closed}
\bibinfo{author}{\bibfnamefont{T.}~\bibnamefont{Dietrich}},
  \bibinfo{author}{\bibfnamefont{S.}~\bibnamefont{Bernuzzi}}, \bibnamefont{and}
  \bibinfo{author}{\bibfnamefont{W.}~\bibnamefont{Tichy}},
  \bibinfo{journal}{Physical Review D} \textbf{\bibinfo{volume}{96}},
  \bibinfo{pages}{121501} (\bibinfo{year}{2017}).

\bibitem[{\citenamefont{Gregory et~al.}(2019)}]{bilby}
\bibinfo{author}{\bibfnamefont{A.}~\bibnamefont{Gregory}} \bibnamefont{et~al.},
  \bibinfo{journal}{Astrophys. J. Supp.} \textbf{\bibinfo{volume}{241}},
  \bibinfo{pages}{27} (\bibinfo{year}{2019}).

\bibitem[{\citenamefont{Romero-Shaw et~al.}(2020)}]{bilby_gwtc1}
\bibinfo{author}{\bibfnamefont{I.~M.} \bibnamefont{Romero-Shaw}}
  \bibnamefont{et~al.} (\bibinfo{year}{2020}),
  \bibinfo{note}{arxiv/2006.00714}.

\bibitem[{\citenamefont{Speagle}(2018)}]{Dynesty}
\bibinfo{author}{\bibfnamefont{J.~S.} \bibnamefont{Speagle}},
  \bibinfo{journal}{MNRAS} \textbf{\bibinfo{volume}{493}},
  \bibinfo{pages}{3132} (\bibinfo{year}{2018}).

\bibitem[{\citenamefont{Skilling}(2004)}]{Skilling}
\bibinfo{author}{\bibfnamefont{J.}~\bibnamefont{Skilling}},
  \bibinfo{journal}{AIP Conf. Proc.} \textbf{\bibinfo{volume}{735}},
  \bibinfo{pages}{395} (\bibinfo{year}{2004}).

\bibitem[{\citenamefont{Landry and Essick}(2019)}]{landryEOS}
\bibinfo{author}{\bibfnamefont{P.}~\bibnamefont{Landry}} \bibnamefont{and}
  \bibinfo{author}{\bibfnamefont{R.}~\bibnamefont{Essick}},
  \bibinfo{journal}{Phys. Rev. D} \textbf{\bibinfo{volume}{99}},
  \bibinfo{pages}{084049} (\bibinfo{year}{2019}).

\bibitem[{\citenamefont{Kish}(1995)}]{Kish}
\bibinfo{author}{\bibfnamefont{L.}~\bibnamefont{Kish}},
  \emph{\bibinfo{title}{Survey Sampling}}
  (\bibinfo{publisher}{Wiley-Interscience}, \bibinfo{address}{Oxford, England},
  \bibinfo{year}{1995}), \bibinfo{edition}{3rd} ed.

\bibitem[{\citenamefont{Víctor~Elvira}(2018)}]{elvira}
\bibinfo{author}{\bibfnamefont{C.~P.~R.} \bibnamefont{Víctor~Elvira},
  \bibfnamefont{Luca~Martino}} (\bibinfo{year}{2018}),
  \bibinfo{note}{arxiv/1809.04129}.

\bibitem[{\citenamefont{Smith et~al.}(2019)\citenamefont{Smith, Ashton,
  Vajpeyi, and Talbot}}]{smith2019expediting}
\bibinfo{author}{\bibfnamefont{R.~J.~E.} \bibnamefont{Smith}},
  \bibinfo{author}{\bibfnamefont{G.}~\bibnamefont{Ashton}},
  \bibinfo{author}{\bibfnamefont{A.}~\bibnamefont{Vajpeyi}}, \bibnamefont{and}
  \bibinfo{author}{\bibfnamefont{C.}~\bibnamefont{Talbot}}
  (\bibinfo{year}{2019}), \eprint{arxiv/1909.11873}.

\bibitem[{\citenamefont{{Lin}}(1991)}]{Lin1991}
\bibinfo{author}{\bibfnamefont{J.}~\bibnamefont{{Lin}}}, \bibinfo{journal}{IEEE
  Transactions on Information Theory} \textbf{\bibinfo{volume}{37}},
  \bibinfo{pages}{145} (\bibinfo{year}{1991}), ISSN \bibinfo{issn}{1557-9654}.

\bibitem[{\citenamefont{Kullback and Leibler}(1951)}]{kullback1951}
\bibinfo{author}{\bibfnamefont{S.}~\bibnamefont{Kullback}} \bibnamefont{and}
  \bibinfo{author}{\bibfnamefont{R.}~\bibnamefont{Leibler}},
  \bibinfo{journal}{Ann. Math. Statist.} \textbf{\bibinfo{volume}{22}},
  \bibinfo{pages}{79} (\bibinfo{year}{1951}).

\bibitem[{\citenamefont{{Veitch} et~al.}(2015)}]{veitch15}
\bibinfo{author}{\bibfnamefont{J.}~\bibnamefont{{Veitch}}}
  \bibnamefont{et~al.}, \bibinfo{journal}{\prd} \textbf{\bibinfo{volume}{91}},
  \bibinfo{eid}{042003} (\bibinfo{year}{2015}).

\bibitem[{\citenamefont{Abbott et~al.}(2019{\natexlab{e}})\citenamefont{Abbott,
   et~al.}}]{abbott2019open}
\bibinfo{author}{\bibfnamefont{R.}~\bibnamefont{Abbott}}, ,
  \bibnamefont{et~al.} (\bibinfo{collaboration}{LIGO Scientific and Virgo
  Collaborations}), \bibinfo{journal}{arxiv/1912.11716}
  (\bibinfo{year}{2019}{\natexlab{e}}).

\bibitem[{psd()}]{psds}
\emph{\bibinfo{title}{Power spectral densities (psd) release for gwtc-1}},
  \urlprefix\url{https://dcc.ligo.org/LIGO-P1900011/public}.

\bibitem[{\citenamefont{Cornish and Littenberg}(2015)}]{cornish2015bayeswave}
\bibinfo{author}{\bibfnamefont{N.~J.} \bibnamefont{Cornish}} \bibnamefont{and}
  \bibinfo{author}{\bibfnamefont{T.~B.} \bibnamefont{Littenberg}},
  \bibinfo{journal}{Classical and Quantum Gravity}
  \textbf{\bibinfo{volume}{32}}, \bibinfo{pages}{135012}
  (\bibinfo{year}{2015}).

\bibitem[{\citenamefont{Littenberg and Cornish}(2015)}]{littenberg2015bayesian}
\bibinfo{author}{\bibfnamefont{T.~B.} \bibnamefont{Littenberg}}
  \bibnamefont{and} \bibinfo{author}{\bibfnamefont{N.~J.}
  \bibnamefont{Cornish}}, \bibinfo{journal}{Physical Review D}
  \textbf{\bibinfo{volume}{91}}, \bibinfo{pages}{084034}
  (\bibinfo{year}{2015}).

\bibitem[{cal()}]{calUncertainties}
\emph{\bibinfo{title}{Calibration uncertainty envelope release for gwtc-1}},
  \urlprefix\url{https://dcc.ligo.org/LIGO-P1900040/public}.

\bibitem[{\citenamefont{Payne et~al.}()\citenamefont{Payne, Talbot, Lasky,
  Thrane, and Kissel}}]{online}
\bibinfo{author}{\bibfnamefont{E.}~\bibnamefont{Payne}},
  \bibinfo{author}{\bibfnamefont{C.}~\bibnamefont{Talbot}},
  \bibinfo{author}{\bibfnamefont{P.}~\bibnamefont{Lasky}},
  \bibinfo{author}{\bibfnamefont{E.}~\bibnamefont{Thrane}}, \bibnamefont{and}
  \bibinfo{author}{\bibfnamefont{J.}~\bibnamefont{Kissel}},
  \urlprefix\url{https://git.ligo.org/ethan.payne/gwtc1_calibration_reweighting}.

\bibitem[{\citenamefont{Vitale et~al.}(2020)\citenamefont{Vitale, Haster, Sun,
  Farr, Goetz, Kissel, and Cahillane}}]{MIT}
\bibinfo{author}{\bibfnamefont{S.}~\bibnamefont{Vitale}},
  \bibinfo{author}{\bibfnamefont{C.-J.} \bibnamefont{Haster}},
  \bibinfo{author}{\bibfnamefont{L.}~\bibnamefont{Sun}},
  \bibinfo{author}{\bibfnamefont{B.}~\bibnamefont{Farr}},
  \bibinfo{author}{\bibfnamefont{E.}~\bibnamefont{Goetz}},
  \bibinfo{author}{\bibfnamefont{J.}~\bibnamefont{Kissel}}, \bibnamefont{and}
  \bibinfo{author}{\bibfnamefont{C.}~\bibnamefont{Cahillane}}
  (\bibinfo{year}{2020}), \bibinfo{note}{dcc/LIGO-P2000293}.

\end{thebibliography}

\end{document}